\documentclass[a4paper,fleqn,usenatbib]{mnras}

\usepackage[utf8]{inputenc}
\usepackage{times}
\usepackage{graphicx,bm,amssymb}
\usepackage{epsfig}
\usepackage{natbib}
\usepackage{psfrag}
\usepackage{graphicx}
\usepackage{ulem}
\usepackage{tikz}
\usepackage{etaremune}
\usepackage{pifont}
\usepackage{colortbl}
\usepackage{orcidlink}
\newcommand{\msun}{\mbox{M$_\odot$}}
\newcommand{\rsun}{\mbox{R$_\odot$}}

\definecolor{ochre}{rgb}{0.8, 0.47, 0.13}

\title[PISN-BBHs from triples with HCA in CEs] {Hypercritical accretion during common envelopes in triples leading to binary black holes in the pair-instability-supernova mass gap}

\date{July 2022}

\author[Moreno M\'endez et al.]
{Enrique Moreno M\'endez$^{1}$\thanks{E-mail: enriquemm@ciencias.unam.mx} \orcidlink{0000-0002-5411-9352}, 
Fabio De Colle$^{2}$
\orcidlink{0000-0002-3137-4633},  
Diego L\'opez-C\'amara$^{3}$
\orcidlink{0000-0001-9512-4177} \newauthor
Alejandro Vigna-G\'omez$^{4}$
\orcidlink{0000-0003-1817-3586}.
\\
$^{1}$Facultad de Ciencias, Universidad Nacional Aut{\'o}noma de M{\'e}xico, A. P. 70-543 04510, CDMX, M\'exico \\
$^{2}$Instituto de Ciencias Nucleares, Universidad Nacional Aut\'onoma de M\'exico, A. P. 70-543 04510 D. F. Mexico \\
$^{3}$C\'atedras CONACyT -- Universidad Nacional Aut\'onoma de M\'exico, Instituto de Astronom\'ia, AP 70-264, CDMX  04510, M\'exico \\
$^{4}$Niels Bohr International Academy, The Niels Bohr Institute, Blegdamsvej 17, DK-2100 Copenhagen, Denmark
}

\begin{document}

\maketitle

\begin{abstract}
Hydrodynamic studies of stellar-mass compact objects (COs) in a common envelope (CE) have shown that the accretion rate onto the CO is a few orders of magnitude below the Bondi-Hoyle-Lyttleton (BHL) estimate. This is several orders of magnitude above the Eddington limit and above the limit for neutrino-cooled accretion (i.e., hypercritical accretion, or HCA). Considering that a binary system inside the CE of a third star accretes material at nearly the same rate as a single object of the same total mass, we propose stellar-evolution channels which form binary black hole (BBH) systems with its component masses within the pair-instability supernova (PISN) mass gap. 
Our model is based on HCA onto the BBH system engulfed into the CE of a massive tertiary star.
Furthermore, we propose a mass transfer mode which allows to store mass lost by the binary onto a third star. Through the use of population synthesis simulations for the evolution of BBHs and standard binary-evolution principles for the interaction with a tertiary star, we are able to produce BBHs masses consistent with those estimated for GW190521. 
We also discuss the massive binary system Mk34 as a possible progenitor of BBHs in the PISN gap, as well as the spin distribution of the observed mergers in the gravitational-wave catalog.
\end{abstract}

\section{Introduction}
\label{sec: intro}

Stellar evolution predicts that low-metallicity stars with initial masses $M_{\rm ZAMS} \lesssim 133 \; \msun$ produce black holes (BHs) with $M_{\rm BH} \lesssim 50 \; \msun$ \citep{2001FWH,heger02,abbott20b}. If a star has $M_{\rm ZAMS} \gtrsim 140 \; \msun$, it may experience a pair-instability supernova \citep[PISN,][]{barkat67,heger02,2003ApJ...591..288H}. PISN occur when a massive oxygen core is sufficiently hot and dense to efficiently create electron-positron pairs. These pairs convert internal gas energy into rest mass, abruptly decreasing the radiation pressure and prompting a rapid contraction of the core. This contraction ignites oxygen, leading to a runaway thermonuclear explosion that leaves no remnant \citep{heger02}. The PISN results in a predicted mass gap for non-interacting massive stars with masses between $50 \; \msun \lesssim M_{\rm gap} \lesssim 130 \; \msun$ \citep[see][and references therein]{2019ApJ...882..121S}.

The binary black hole (BBH) system progenitor of the GW source GW190521 consisted of two black holes (BHs) with masses $M_{\rm BH, 1} = 85^{+21}_{-14}\msun$ and $M_{\rm BH, 2} = 66 ^{+17}_{-18}\msun$ \citep{abbott20a} which fall in the theoretically predicted PISN mass gap. The merger of the BBH radiated about $7.6^{+2.2}_{-1.9} \; \msun$ in GWs, and resulted in a single 142$^{+28}_{-16}\msun$ BH \citep{abbott20a}. The inferred Kerr spin parameters of the merging BHs was $0.1 \lesssim a_{\star} \lesssim 0.9$, where $a_\star \equiv Jc^2/GM_{\rm BH}^2$, being $J$ the angular momentum of the BH, and the spin parameter of the remnant BH was $a_\star \simeq 0.72^{+0.09}_{-0.12}$. 

Several mechanisms have been proposed to produce BBH in the PISN gap. These include: {\it a)} multiple generations BBH mergers either in clusters  \citep[e.g.,][]{fragione20, liulai20, stone19}, in AGN disks \citep{mcKernan12, bartos17}, or in the field \citep{VignaGomez2021}; {\it b)} accretion in proto-clusters \citep{roupaskazanas19a, roupaskazanas19b}, isolated binaries \citep{vanson20}, halos \citep{safarzadeh20}, or onto primordial BHs \citep{deluca20}; {\it c)} creation of massive BBHs that form from isolated binaries \citep{spera19}, the merger of two stars below the PISN forming a star with a small core and an oversized envelope \citep{renzo20}, massive BHs from population III stars \citep{kinugawa20, vink21};  {\it d)} mechanisms beyond the standard models of stellar evolution \citep[]{sakstein20, straight20, ziegler20}.

The common envelope (CE) phase is a short lived stage during the evolution of a binary system in which the envelope of one of the stellar components engulfs the other component \citep{paczynski1976, ibenlivio1993}. Due to mass transfer and orbital decay the orbital separation decreases up to a point in which highly energetic and transitory phenomena may be produced, for example: type Ia supernovae \citep[][]{chevalier2012, py2014}, short GRBs \citep[][]{vigna2020}, long GRBs \citep[][]{brown2007,2011MM}, and GWs \citep[][]{abbott2016c}. During the CE phase, which may occur in several occasions during the evolution of the binary system, compact objects may be the engulfed component \citep{ivanova13} and the accretion onto the secondary component may play an important role \citep{chamandy18a}.

Most studies (e.g., those in the third paragraph) assume that accretion is limited by the Eddington luminosity. However, accretion onto a single star or a binary system may be super-Eddington or hypercritical\footnote{In super-Eddington accretion ($\dot{M}_{\rm Edd} < \dot{M} \lesssim 3000\;  \dot{M}_{\rm Edd}$) the energy excess is lost through photons, while in hypercritical accretion (HCA) the thermal energy is mainly lost through neutrinos, given that for $\dot{M} \gtrsim 3000 \dot{M}_{\rm Edd}$ and the accreting material has large density and temperature \citep{1989Chevalier,BrownW1994}.} \citep[e.g.,][]{pacucci15, sakurai16, sugimura17, li19}. This is especially true for a compact object (CO) accretor. Radiation-hydrodynamics simulations have shown that steady accretion happens once the accretion rate is $\dot{M} \gtrsim 3000\; \dot{M}_{\rm Edd}$, with $\dot{M}_{\rm Edd}$ being the Eddington-limited mass-accretion rate \citep{inayoshi16}. In addition, HCA can also produce the masses and spins observed in some BH binaries in high-mass, X-ray binary systems \citep{2011MMBLW,2014MM}.

In this paper, we propose that HCA can play an important role in the formation of massive stellar BHs. We show that a BBH system with HCA during a CE phase may produce BHs with masses within the PISN gap. We discuss two possible scenarios: a) the evolution of a hierarchical triple system, or b) the outcome of a BBH capturing a massive star. We propose a mechanism which allows the system to {\it store} material, initially part of the more massive binary stars, in the tertiary star. Both scenarios produce BBH with component masses within the PISN mass gap through HCA during the CE phase of the massive star.

The paper is organized as follows. In Section~\ref{sec:HCA} we discuss the conditions for HCA onto a BBH during the CE phase. In Section ~\ref{sec:MTnMS} we discuss mass-transfer and storage in triple systems. In Section~\ref{sec:scenarios} we describe possible evolutionary paths of triple-systems which produce hyper accreting BHs within the CE of a red giant or red super giant. In Section~\ref{sec:disc} and ~\ref{sec:concl} we discuss our methods and results and drive our conclusions.

\section{Hypercritical accretion onto a BBH system during the CE phase}
\label{sec:HCA}

In this section we describe how during HCA onto a BBH system which is within the CE of a massive star a non-negligible fraction of the massive star may be accreted by the BBH. We also discuss how accretion and the ejection process are the key competing mechanisms to determine the amount of hydrogen-envelope accreted onto the inner BBH. 

\subsection{Accretion onto a BBH}
\label{subsec:AccRate}

Previous studies have shown that accretion onto a star or compact object orbiting inside the CE of a red giant (RG) or red super-giant (RSG) occurs at a fraction ($\epsilon \approx 0.01-0.1$) of the Bondi-Hoyle-Littleton rate \citep[see, e.g.,][]{ricker12, macleod15, MM17,chamandy18a,LC19,LC20}. The accretion rate onto the companion ($\dot{M}_{\rm a}$) is:
\begin{eqnarray}
    \dot{M}_{\rm a} &\equiv& \epsilon \dot{M}_{\rm BHL} =\epsilon \frac{4 \pi \rho G^2 M^2}{(v^2+c_s^2)^{3/2}} \nonumber \\
    &\simeq& 2.8\times10^{-5} \ \epsilon_{0.1} \ \rho_{-6} \ M_{50}^2\ v_7^{-3} \ \msun \; {\rm s}^{-1}
  \label{eq:BHL}
\end{eqnarray}
where $G$ is the Gravitational constant, $\rho_{-6}$ is the density of the environment in units of $10^{-6}$g~cm$^{-3}$, $M_{50}$ the mass of the accretor in units of 50 $\msun$, $v_{7}$ the velocity of the accretor in units of $10^{7}$cm~s$^{-1}$, and $\epsilon_{0.1}$ the accretion efficiency normalized to $\epsilon =0.1$ (which depends on the structure of the envelope, in particular on the presence of density and velocity gradients). Since the sound speed within the envelope of a RG or RSG is $c_s \sim 10\; (T/10^4 {\rm K})^{1/2}$ km s$^{-1}$, and the velocity of the accreting material is of order of the keplerian velocity ($v \sim v_{k}\gtrsim 100$ km s$^{-1}$), we have $(v^2+c_s^2)^{3/2}\sim v^3$.

If the secondary is engulfed by the CE in a sudden plunge, the orbital period of the accretor as it enters the envelope of the CE may be completely asynchronised with respect to the rotation period of the donor. This scenario leads, at first order approximation, to BHL accretion. The case of a gradual plunge occurs when the donor and the accretor are fully synchronized, and thus, the relative velocity is zero. On the other hand, if the secondary is engulfed gradually by the CE, then approximately, Bondi accretion takes place (case in which $v\lesssim c_s$ in equation \ref{eq:BHL}). Density and velocity gradients, magnetic fields, convection flows, Coriolis, centrifugal and Euler forces, and the presence of a preexisting disk, to name some, modify the simple picture of BHL or Bondi accretion, and thus it is uncertain, in any case, whether a disk may or may not form around the accretor. 

Meanwhile, accretion onto a BBH system depends on the orbital separation $a$ between the BHs and on the accretion radius ($r_a = 2 G M/v^2 \sim 3 \times 10^{14}~M_{100}~v_7^{-2}~{\rm cm}$; see, e.g., \citealt{edgar04}). If the BBH forms part of a triple system, a fraction of the envelope of the tertiary can be accreted during a CE phase by the CE-engulfed BBH (see Section~\ref{subsec:Stability}). Also, in the case of a triple system, it must be considered that the Keplerian velocity of the binary will never be zero, regardless of whether synchronization is achieved or not with the envelope of the tertiary.

In order to model the effect of accretion of a binary system inside a CE of a third star, we follow \citet{comerford19, comerford20}. These studies present numerical simulations of accretion onto binaries for a wide range of orbital periods and inclinations. They find that for small orbital separations ($a \ll r_a$) the accretion takes place onto a point mass (located at the center of mass of the BBH) with a mass given by the sum of the two individual masses ($M=M_{\rm BH_1} + M_{\rm BH_2}$). When $a \gg r_a$, each BH accretes as a single object. The intermediate case ($a \sim r_a$) is complex and remains to be fully studied. In the following we will focus on the case in which $a\gg r_a$.

From equation \ref{eq:BHL} it is clear that the mass of the black holes ($M_{\rm BH,i}$, with $i=1,2$) changes as $\dot{M}_{\rm BH,i} = k M^2_{\rm BH,i}$ (where $k= 4 \pi \epsilon \rho G^2/v^3$), which by direct integration gives:
\begin{equation}
  M_{\rm BH,i} = \frac{M_{{\rm BH,{i_0}}}}{1- k M_{{\rm BH,{i_0}}}t}\;,
\end{equation}
where $M_{{\rm BH,{i_0}}}$ is the initial mass of the black hole $i$. The black hole $i$ achieves a mass $M_{\rm BH,i}$ in a time:
\begin{equation}
  t_{\rm f,i} = \left( 1-\frac{M_{{\rm BH,{i_0}}}}{M_{\rm BH,i}}\right) \frac{1}{k M_{{\rm BH,{i_0}}}}\;.
\end{equation}
Since $t_{\rm f,1} = t_{\rm f,2}$, then, the following relationship between the ratio of the initial and the final masses of the black holes ($q_0=M_{{\rm BH,{2_0}}}/M_{{\rm BH,{1_0}}}$ and $q=M_{\rm BH,2}/M_{\rm BH,1}$ respectively) is obtained:
\begin{equation}
   q_0 = \frac{q M_{\rm BH,1}/M_{{\rm BH,{1_0}}}}{1+q( M_{\rm BH,1}/M_{{\rm BH,{1_0}}}-1)}\;.
\end{equation}
Using equations 2, 3, and 4 the initial masses of the BHs and the accretion time $t_f$ (for the BHs to reach their final masses), are obtained as a function of the final masses.

\begin{figure}
\centering
\includegraphics[scale=0.35]{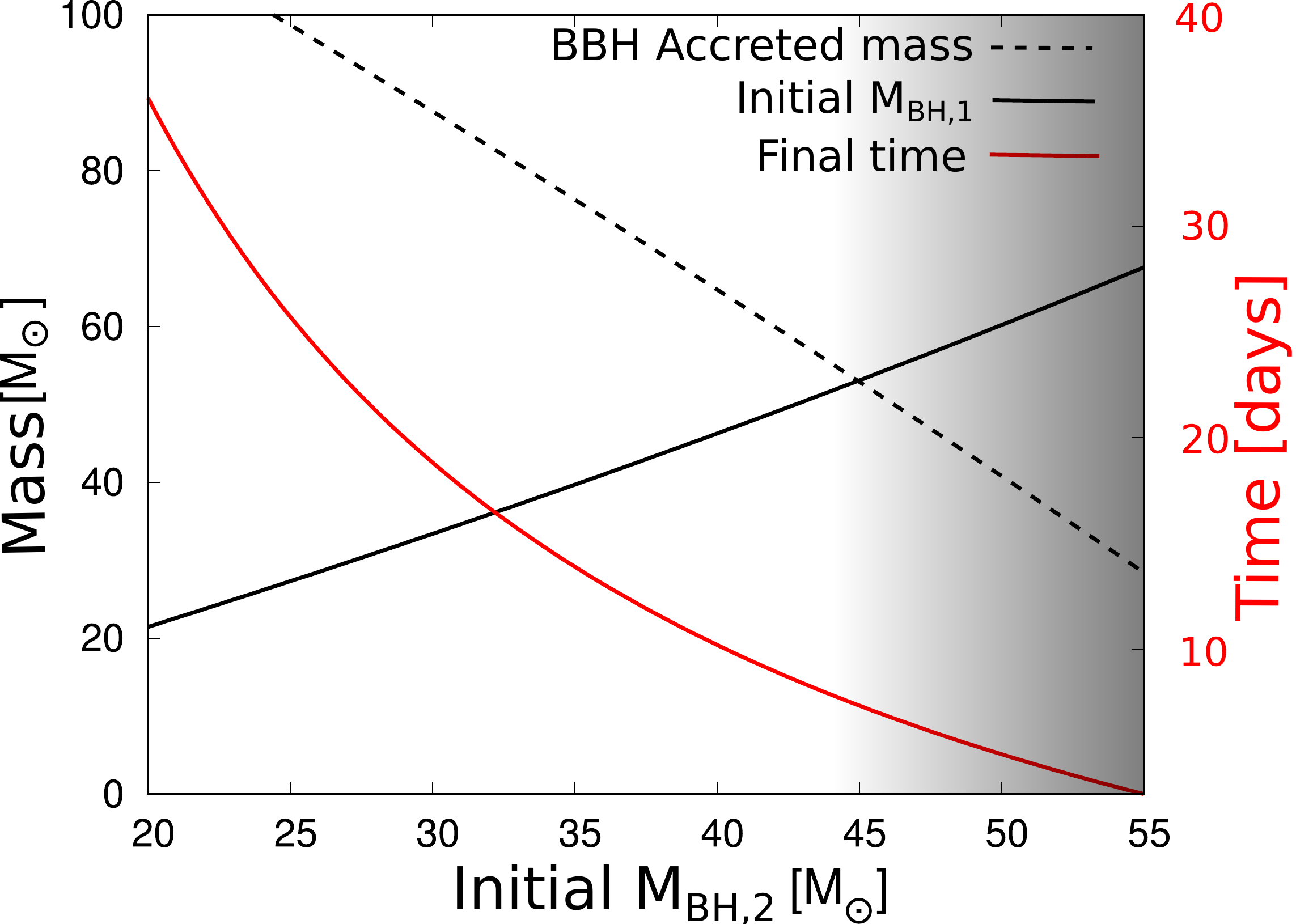}
\caption{The solid black line indicates the initial mass of the primary BH ($M_{\rm BH,1}$) as a function of the initial mass of the secondary BH ($M_{\rm BH,2}$) in the binary system. The dashed black line represents the mass accreted by the BBH system. The red solid line and the right $y$-axis show the accretion time needed for the primary and secondary BHs to end up with 85 $\msun$ and 65 $\msun$ respectively during the CE phase (assuming an accretion rate given by 10\% of the BHL rate). The gray shaded area is excluded as the initial mass of the primary BH is larger than $\gtrsim$ 55 $\msun$. Larger BH masses are excluded by single stellar evolution models (see Section~\ref{sec: intro}).}
\label{fig3}
\end{figure}

In Figure~\ref{fig3} we plot the evolution of the BBH during HCA when inside the CE of a third star. Specifically, we plot the range of possible initial masses of the two merging BHs (with $M_{\rm BH,1} > M_{\rm BH,2}$), the amount of mass accreted by the binary system, and the time required for the BHs to reach the masses of GW190521 (85$\msun$ and 66$\msun$). For the latter, we assume that the accretion rate onto the BBH during the CE phase is $10\%$ of the BHL accretion rate. The accreted mass will be a substantial fraction of the envelope of the tertiary star in which the BBH is immerse. If the initial mass of both BHs is large, the accretion time required to reach masses similar to those of GW190521 is considerably smaller (a few days to over a month) and the required mass transferred and accreted from the donor star drops to a few tens of solar masses. For example, if the initial mass of the secondary BH is $M_{\rm BH,2}\approx 40\; \msun$, then the mass of the primary BH must be $M_{\rm BH,1}\approx 45\; \msun$ (from the solid black line) and the accretion time for each of the BHs to reach $65\; \msun$ and $85\; \msun$ (respectively, and assuming 10\% of the BHL accretion rate) is $\simeq 10$ d (red solid line). The total mass transferred and accreted from the envelope of the tertiary star needs to be $\Delta M_{\rm CE}$ $\gtrsim 65\; \msun$ (dashed black line). Assuming a maximum available envelope mass of 100 $\msun$ then $\gtrsim 65\%$ of the envelope must be accreted (for further details see Section~\ref{sec:scenarios}).

\subsection{Limits on the accretion onto the BBH}
\label{subsec:Stability}

The accretion onto a BBH system may be limited by several effects: a) the radiation pressure of the accreting material, b) the orbital timescale, and c) the amount of material ejected from the CE. In this section, we discuss these effects and argue that they do not represent a strong limiting factor to the accretion onto the BBH.

Firstly, we discuss the feedback due to radiation pressure and neutrino effects. The accretion rate given by equation ($\ref{eq:BHL}$) is $\sim$10 orders of magnitude above the Eddington mass accretion rate given by:
\begin{equation}\label{eq:EddLim}
    \dot{M}_{\rm Edd} = \frac{L_{\rm Edd}}{c^2} \simeq 3.5\times 10^{-15} M_{50}\ \msun \; {\rm s}^{-1}\;,
\end{equation}
where $L_{\rm Edd}$ is the Eddington luminosity and $c$ is the speed of light. Nevertheless, the Eddington limit assumes spherical symmetry, i.e., Bondi accretion. BHL accretion breaks spherical symmetry, leading to radiation ejected through polar jets and material accreted via the equatorial plane. Additionally, velocity and density gradients, as well as Coriolis and centrifugal forces, may further break the cylindrical symmetry of the BHL accretion \citep{MM17,LC19,LC20}.

Thus, the BHL accretion rate onto a BH which is within the CE of a massive RG or RSG is HCA. In this case, as the accretion rate is $\dot{M}_{\rm a}\gtrsim 3\times 10^3 \; \dot{M}_{\rm Edd}$, neutrinos are produced \citep{HouckCh1991,BrownW1994}. Since the density and temperature of the material in the accretion flows around the BH are well below those in core-collapse supernovae, close to where the Eddington limit for neutrinos lies, these neutrinos will not interact with the accretion flow. Therefore, most of the released binding energy will be radiated away by neutrinos which leave the CE without exerting pressure on the accreting material and allowing for extremely large accretion rates. Hence, the radiation pressure and neutrino emissivity will not have effect on our results.

Second, we consider the limit imposed on accretion due to the orbital timescale. With $\dot{M}_{\rm a}\gtrsim 10^{9} \dot{M}_{\rm Edd}$ (see equations 1 and 5), and most of the binding energy being radiated away by neutrinos, the BBH could increase its mass from $M_{\rm BBH,i} \simeq 55 \;\msun + 45\; \msun = 100\; \msun$ to $M_{\rm BBH,f} \simeq 85\; \msun + 65\;\msun = 150\; \msun$ within a few weeks of CE (see Figure~\ref{fig3}). However, if the initial orbital separation in the triple system is of order $a_{\rm tri,i} \gtrsim 10^{14}$ cm (considering a $M_\star  = 150 \msun$ donor star as well as the BBH system), the initial orbital period will be:
    \begin{equation}
        T = \frac{2\pi R}{v_k} = 2\pi \sqrt{\frac{R^3}{GM}} \sim  100~a_{\rm tri,14}^{3/2}~M_{100}^{-1/2}~{\rm d}\;.
    \end{equation}

As the BBH can not accrete an amount of mass larger than the stellar material available during their orbit around the companion star, the timescale for the accretion onto the BBH system will be the largest between $T$ and $\tau_{\rm BHL}\sim 30 \; (\epsilon/0.1)^{-1}$ d. Since the BHL accretion timescale ($\tau_{\rm {BHL}}$) is smaller than the orbital timescale, accretion is limited by $T$. As the orbital separation decreases, so does the orbital period, and it continues to decrease until $T\sim \tau_{\rm {BHL}}$.

Last, the ejection of stellar material from the CE could limit accretion onto the BBH (as the material ejected from the CE and that lost by the triple system could unbinding the envelope). The BBH mass\footnote{Here, by accretion on the BBH we refer to the accretion onto the system which could include an accretion disk around the BBH. The actual accretion onto the BBH will be then delayed by the viscous time.} increases as {$\dot{M}_a = \epsilon \dot{M}_{\rm BHL}$ (equation \ref{eq:BHL})}, while the envelope mass drops by the same amount, i.e. $\dot{M}_{\rm env} = - \dot{M}_a$. We limit the amount of accreted mass as the minimum between the BHL accreted one and the mass available at radius larger than the orbital separation ($M_\star(r>a)$), that is:
\begin{eqnarray}
    M_{\rm BBH}(t) &=& M_{\rm BBH, 0} + \min\left(M_*(r>a),\int_0^t \dot{M}_a(t') dt'\right) \;, 
        \label{eq:menv} \\
    M_{\rm env}(t) &=& M_{\rm env, 0} - (M_{\rm BBH}(t) - M_{\rm BBH, 0})\:,
\end{eqnarray}
where $M_{\rm BBH, 0}$, $M_{\rm env, 0}$ are the initial BBH and envelope masses.

The inspiral of the BBH will be driven by the the gravitational drag due to the envelope material accreting onto the BBH \citep{livio88}, which produces a change in the BBH orbital energy (around the massive star) as:
\begin{equation}
    \frac{dE_{\rm orb}}{dt} = - \dot{M}_a v_k^2\;,
\label{eq:dedt}
\end{equation}
where $\dot{M}_a$ is the mass accretion rate (given by equation \ref{eq:BHL}), $v_k$ is the Keplerian velocity ($v_k=\sqrt{G M_\star(r \leq a)/a}$, with $M_\star(r \leq a)$ the enclosed mass within the orbital separation), and the orbital energy is $E_{\rm orb} = -0.5\;G M_{\rm BBH} M_\star(r \leq a)/a$. Assuming that a fraction $\alpha$ of the orbital energy is deposited (as thermal energy) into the envelope, thus, the energy deposited ($E_d$) is given by
\begin{equation}
    \frac{dE_d}{dt} = \alpha \dot{M}_a v_k^2\;,
\label{eq:deddt}
\end{equation}

Taking the $\alpha\lambda$ energy formalism of \citet{vdheuvel1976} in which the envelope will be ejected once the energy deposited is of order of the binding energy \citep[e.g.,][]{webbink84, olejak21}, and using equations \ref{eq:dedt}, \ref{eq:deddt} (as well as the binding energy of the envelope, $E_b$), then the condition $E_b\sim E_d$ reads
\begin{equation}
  \frac{G M(a)M(r>a_f)}{\alpha \lambda a}
\sim \int_0^t \dot{M}_a v_k^2 dt\;,
\label{eq:alphalambda}  
\end{equation}
where $\lambda$ is a numerical factor which depends on the  density structure of the star \citep{1990ApJ...358..189D}.

By numerically integrating this set of equations (\ref{eq:menv}-\ref{eq:alphalambda}) (and considering a stellar core with a mass $35\ \msun$, an envelope density profile, between $2\times 10^{11}$ cm and $2 \times 10^{13}$ cm, $\rho \propto r^{-n}$ (with $n=2.5-2.8$) and a total envelope mass of  65 $\msun$, the BBH inspiral time is $\sim 100-170$~days, and most the material can be accreted by the BBH (since $E_b\gtrsim 10 E_d$ at all times) for $\alpha \leq 1$. Thus, the loss of material during the accretion phase is minor (at least, in the framework of this simple analytical method), and the system is stable.

\subsection{Jets as an extra energy injection source.}
\label{subsec:Jets}

If an accretion disk does not form around the accreting component, HCA may only proceed through neutrino-radiation cooling. \citep[i.e., the accretion rate must be at least three orders of magnitude above the Eddington limit; a condition which we have shown to be vastly fulfilled by our models;][]{1989Chevalier,BrownW1994}. In this case, there is no extra energy injection into the CE other than the orbital energy, parametrised by the $\alpha\lambda$ formalism. On the other hand, when an accretion disk forms around an accreting BH, a collimated and relativistic jet may be launched \citep{armitagelivio2000, soker2004}. In this case, as shown by \citet{MM17}, the energy deposition into the CE will dominate over the $\alpha\lambda$ and other energy sources \citep[e.g., recombination energy, enthalpy, tidal heating, magnetic fields, see][for further details]{ivanova13}. Switching on this mechanism can bring the system into the so-called grazing envelope phase \citep{ss2015, soker2015, sks2017, ss2018} and may even terminate the CE phase in extreme cases by removing the envelope \citep{soker2017}. Nonetheless, as the energy deposited by the jet will drop the accretion rate, the jet power will also drop, then alternating periods of accretion with period of quiescence. The ``self-regulation'' of the jet guarantees that a significant percentage of the CE available material will be accreted (although it does not guarantee that the orbital separation will decrease). Given the complexity of the problem, future numerical works are needed to understand the full impact of jets or outflows in the CE evolution (hereafter, CEE).  

\section{Mass-transfer and mass-storage in triple systems}
\label{sec:MTnMS}

One of the main processes determining the evolution of a multiple-stellar system is mass transfer between the components. In this Section, we describe our assumptions and the uncertainties on mass transfer mechanism, particularly in the context of a CEE from a tertiary donor with an inner BBH accretor. We provide a qualitative description of each mechanism and we estimate the physical quantities (in particular, orbital separations and periods) characterizing the system.

In the Roche-lobe overflow (RLOF) mechanism the mass is transferred to the companion once the radius of the envelope of the donor exceeds the volume-averaged Roche-lobe (RL) radius $r_{\rm RL}$ \citep{eggleton83}:
    \begin{equation}
       \frac{r_{\rm RL}}{a} = \frac{0.49 q^{2/3}}{0.6 q^{2/3} + \ln(1 + q^{1/3})} \; ,
    \label{eq:RL}
    \end{equation}
where $q$ is the mass ratio of the binary components ($q=M_{\rm{d}}/M_{\rm{a}}$, with $M_{\rm{d}}$ the donor and $M_{\rm{a}}$ the accretor).

During RLOF two important processes may allow a binary system to transfer mass \citep[][]{1994IntBins}, these are: a) when the donor star is more massive than the accretor the orbital separation decreases (as well as the RL of the donor); b) the reaction of the donor star to the loss of its outer layers. In the latter if the envelope is radiative it will contract as a response to mass loss, while if it is convective it will expand \citep[e.g.,][]{soberman97}. The envelopes of RG stars is usually radiative, however, RSG stars have convective envelopes \citep{klencki20}. If the star expands, it is likely that the star will continue to fill its RL (or that it will fill it up again, if the orbital separation increased); this process occurs in a Kelvin-Helmholtz (or thermal) timescale \citep{1994IntBins}. In order to maintain a stable RLOF phase, at least one of these runaway processes must be at play \citep{soberman97}.

Another important mass-transfer mechanism in the evolution of multiple stellar systems is the wind RLOF (wRLOF), in which the wind escaping the donor star is swept by the accreting object. For close binary systems, \citet{mohamed07} found that mass transfer through wRLOF can be highly efficient (with up to $\sim 70\%$ of the wind being captured and accreted by the companion), when the orbital velocity of the binary is comparable to the wind velocity. A compact orbit is a necessary condition for a highly efficient wRLOF mass-transfer, because it allows for a large orbital velocity as well as a stellar wind which has not been fully accelerated. Stellar winds, particularly in massive stars, require $\sim 10-15\ \rsun$ to fully accelerate, hence the winds in this scenario have not reached escape velocity and can be substantially funneled through the first Lagrange point (L1) and focused towards the accretor \citep[e.g.,][]{vink11}. This mechanism is an efficient mass transfer mechanism in the evolutionary stages of the massive stellar systems \citep{2018Vink}.

An additional mass-transfer mechanism may occur in triple systems, specifically, when the inner binary system is in a CE phase and the tertiary is far orbiting around the binary system. In this stage, a substantial amount of material from the outer layers of a CE may be inflated without being fully unbound from the binary system; in fact, this material may even form a circumbinary disk \citep{2011KashiSoker}. This material may slowly move to a large orbit around the binary (and may even fall back towards the binary if its velocity is below the escape velocity). So, if a tertiary star orbits the binary it may be able to capture a large fraction of this material, see Section~\ref{subsec:triple} and Section~\ref{subsec:ce-bbh-capture} for further details. This mass-transfer mechanism can be important in triple systems and we term it {\it common-envelope Roche-lobe overflow} (ceRLOF).

In principle, ceRLOF allows for an effective mechanism to transfer mass in a conservative way (hereafter, we refer to conservative mass-transfer as a process in which the mass lost from the binary system is accreted by the tertiary star) and store a substantial fraction of the envelopes of both stars in the inner binary in the tertiary star. During the evolution of our triple-stellar systems, mass is transferred back and forth between the stellar components at different stages, particularly during CE stages from the inner binary (ceRLOF). The captured mass will later be returned to the inner BBH after the tertiary, and last-to-evolve star, becomes a giant which fills its RL and engulfs the BBH in a third CE stage. In this scenario, a large-mass BBH can be formed successfully (see Section~\ref{sec:HCA}) without requiring extremely massive stars ($M_\star > 150 \msun$) or systems with more than three stars\footnote{We assume that the two SNe forming heavy BHs lose only a small fraction of the mass of binary \citep[for further details see][]{fryer2012}.}.

\section{CE-BBH scenarios}
\label{sec:scenarios}

In this section, we discuss two possible scenarios which may produce a BBH system with masses close to those in the PISN mass gap which have been inferred for GW190521 by the Ligo-Virgo-Kagra collaboration. First, we describe the evolution of a binary system, composed by two massive stars, which produces a BBH (Section~\ref{subsec:inner_binary}). Then, we consider the case of ceRLOF from a tertiary star in the case when the tertiary is pre-existent (Section~\ref{subsec:triple}), and the case in which a tertiary star has been captured by the BBH system (Section~\ref{subsec:ce-bbh-capture}).

\subsection{Evolution of the inner massive binary in a hierarchical triple stellar system}
\label{subsec:inner_binary}

In order to look for a binary a binary system with component masses close to the lower edge of the pair-instability black-hole mass gap ($M_{\rm BH,1} \sim M_{\rm BH,2} \sim 45\msun$) we use the population synthesis code COMPAS \citep[see][and references therein]{riley2022}.We adopt the default choices of COMPAS parameters for binary evolution which follows the classic CE formation scenario of BBH mergers \citep[e.g.,][]{2016Natur.534..512B,2017NatCo...814906S,2018MNRAS.481.4009V}. The evolution of such binary is shown in Figure~\ref{BE_CEE} and discussed next (see also Table~\ref{tab1}):

\begin{figure}
\centering
\includegraphics[width=\columnwidth]{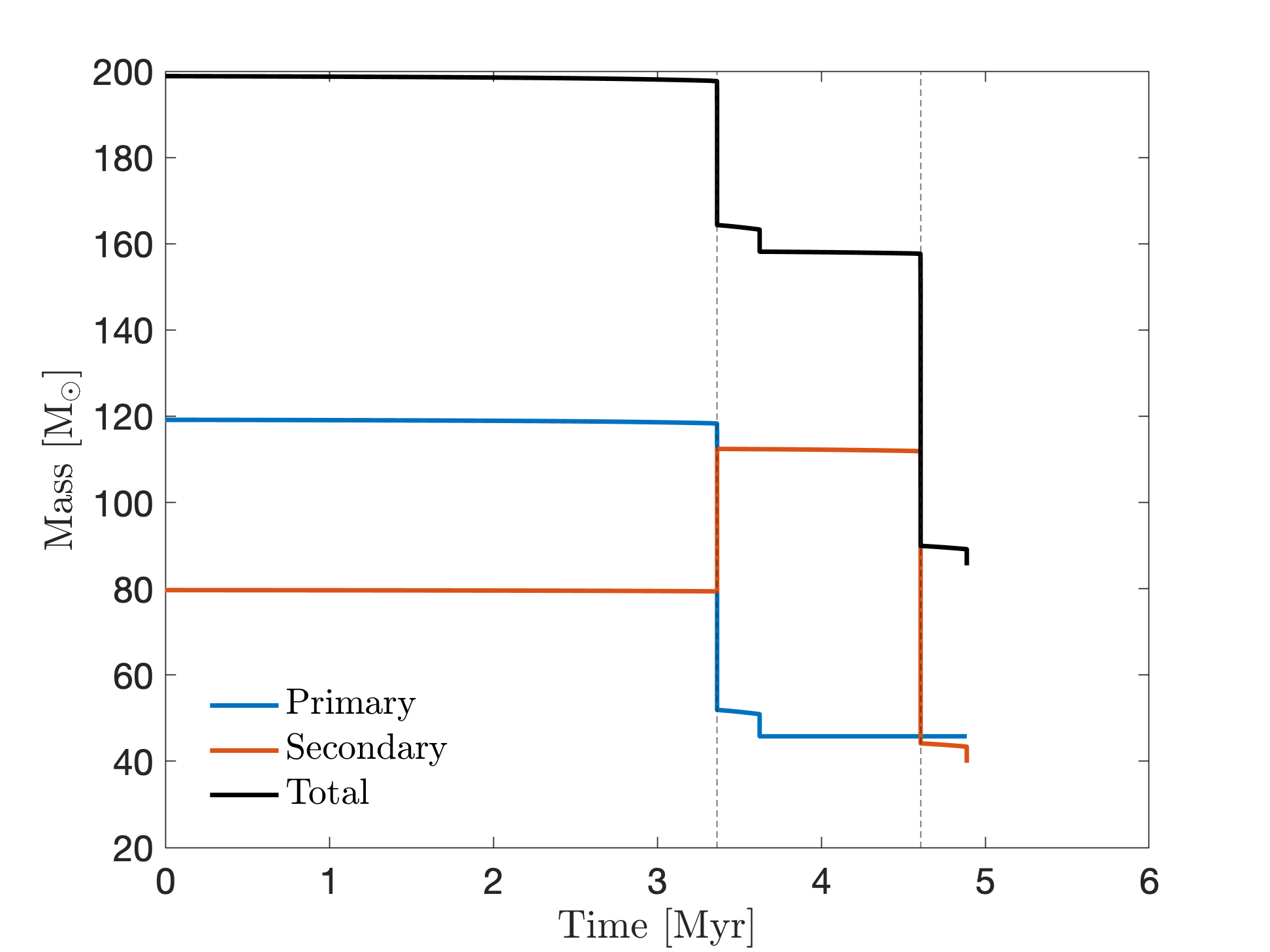}
\includegraphics[width=\columnwidth]{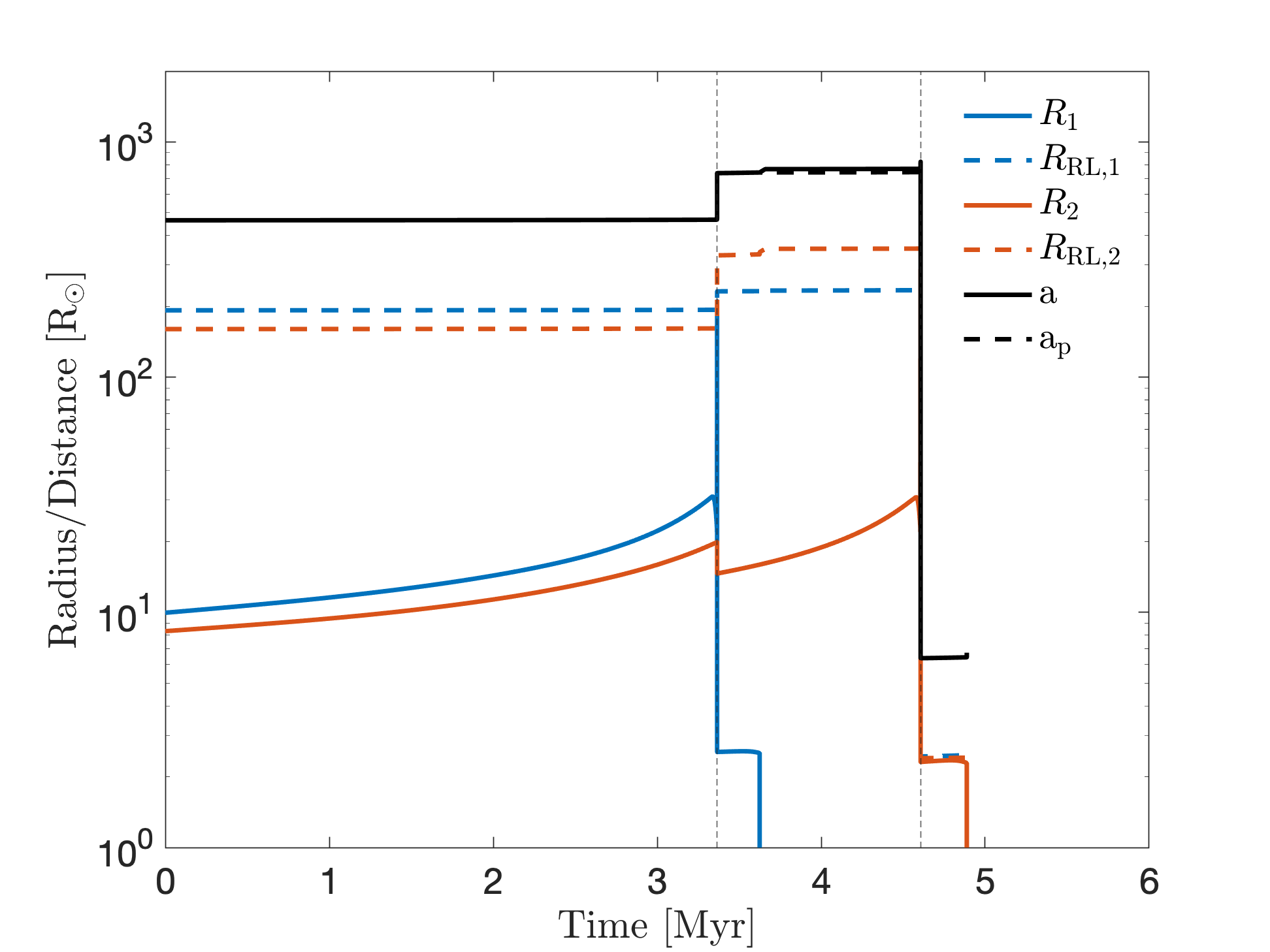}
\caption{Evolutionary pathway for a binary system with a primary star with $M_1=119.2\ \msun$, secondary with $M_2=79.7\ \msun$ (both with $Z=0.0001$), and an orbital separation of $a=464.6\ \rsun$. The upper panel shows the evolution of the primary, secondary and the total masses. The bottom panel shows the evolution of the radii of the two stars ($R_1$ and $R_2$), the correspondent Roche lobe radii of each of the stars ($R_{\rm RL,1}$ and $R_{\rm RL,2}$), and the orbital separation of the binary (semi-major axis $a$ and periastron $a_{\rm p}$).}
\label{BE_CEE}
\end{figure}

\begin{table*}{}
\centering
\begin{tabular}{ |ll|cccccccccc| } 
\hline\hline
Stage & Event & Time & M$_{1}$ & M$_{2}$ & M$_{3}$ & $a_{\rm{in}}$ & $a_{\rm{out,crit}}$ & $a_{\rm{out}}$ & $R_{\rm{RL,12}}$& $R_{\rm{RL,3}}$ & $\Delta M$ \\
& & [Myr] & [$\msun$] & [$\msun$] & [$\msun$] 
& [$\rsun$] & [$\rsun$] & [$\rsun$] & [$\rsun$] & [$\rsun$] & [$\msun$]\\
\hline\hline
i   & ZAMS     & 0.0 & 119.2 & 79.7  & 35  & 464.6  & $>3 a_{\rm{in}}$ & 1817 & 966  & 441 & 0.0 \\ 
ii  & RLOF     & 3.4 & 51.9  & 112.4 & 35  & 737.1  & $>3 a_{\rm{in}}$ & 2216 & 1142 & 568 & 33.4 \\ 
iii & SN       & 3.6 & 45.8  & 112.4 & 35  & 766.5  & $3 a_{\rm{in}}$  & 2300 & 1178 & 595 & 0.0 \\
iv  & CEE      & 4.6 & 45.8  & 44.1  & 103 & 6.4    & $>3 a_{\rm{in}}$ & 818  & 300  & 320 & 68.0 \\
v   & SN       & 4.9 & 45.8  & 39.6  & 103 & 6.7    & -                & 836  & 303  & 330 & 0.0 \\
vi  & RLOF/CEE & 5.0 & 86.0    & 66.0  & 37  & $<6.7$ & -                & 2074 & 1045 & 551 & (40 + 26)* \\
\hline\hline
\end{tabular}
\caption{Principal stages (and main values) of the evolution of the binary shown in Figure~\ref{BE_CEE}. *Note: $40~\msun$ are accreted by the most massive BH ($M_1$) and $26~\msun$ by the least massive BH ($M_2$).}
\label{tab1}
\end{table*}

\begin{enumerate}[i]
    \item The binary system begins at the zero-age main sequence (ZAMS) with a primary of 119.2 $\msun$ and a companion of 79.7 $\msun$ (both with a metallicity of $Z=0.0001$) in a circular orbit with a semi-major axis of 464.6 $\rsun$.
    \item At $\approx$ 3.4 Myr, the post-main-sequence primary fills its RL and begins a semi-conservative episode of mass transfer with the main-sequence companion.
    In this first episode, the primary donates $\approx$ 67.1 $\msun$, of which, $\approx$ 33.0 $\msun$ are accreted by the main-sequence companion, and the remaining $\approx$ 34.1 $\msun$ are lost from the (inner) binary.
    This semi-conservative episode where mass is lost via isotropic re-emission widens the (inner) orbit to $\approx$ 737.1 $\rsun$.
    \item Shortly after, at $\approx$ 3.6 Myr, the primary collapses into a 45.8 $\msun$ BH.
    \item The secondary star finishes the main sequence and enters the Hertzsprung gap. At $\approx$ 4.6 Myr it begins a dynamically unstable mass transfer episode.  This mass transfer episode results in a common-envelope episode, a process where $\approx$ 68 $\msun$ of envelope can be removed from the (inner) binary. This process results in an $\approx$ 44.1 $\msun$ stripped star in a close binary with orbital separation of several solar radii.
    \item Finally, the secondary collapses at $\approx$ 4.9 Myr to form an $\approx$ 39.6 $\msun$ BH.
\end{enumerate}
We choose this low-metallicity binary system in order to neglect the effect of stellar winds in our calculations. Moreover, lower metallicity environments are predicted to lead to more massive BBH remnants \cite[e.g.,][]{2019MNRAS.490.3740N} and compact BBHs \citep[e.g.,][]{VignaGomez2021}.

\subsection{Hierarchical triple system}
\label{subsec:triple}

The hierarchical-triple-system (HTS) channel which we propose involves a massive inner stellar binary (described in Section~\ref{subsec:inner_binary}) and a dynamically-stable tertiary star orbiting around it \citep[e.g.,][]{hutbahcall83}. The tertiary can capture and store a large fraction of the mass which is ejected from the inner binary system as it becomes a BBH system. Once the BBH is produced the tertiary may begin its RLOF phase and with this transfer mass back onto the BBH. Such a system may form in a massive stellar-forming regions via gravitational capture, or as a triple system in the field \citep{2017ApJS..230...15M}.

The condition for dynamical stability of a HTS is that the orbit of the tertiary star ($a_{\rm out}$) is always larger than three times the orbit of the binary ($a_{\rm in}$), this is, $a_{\rm out} \gtrsim 3 a_{\rm in}$ \citep{kozai1962, lidov1962, shevchenko2017}. To simplify our model, we assume that mass is transferred from the inner binary and onto the tertiary star during the ceRLOF stage (and not during the RLOF stage when the primary star expands; nonetheless this early stage could also transfer up to 33.4$\msun$ onto the tertiary). If this was not the case, the initial orbital separation of the tertiary star, as well as the subsequent stages, would need to be larger to prevent the HTS from becoming unstable.

Using the results from COMPAS for the binary system (see Section~\ref{subsec:inner_binary}), and for the tertiary star evolution either conservative mass transfer during RLOF, CEE, or ceRLOF stages, as well as isotropic mass loss during SNe, we broadly calculate the evolution of the HTS (see Table~\ref{tab1} for details). As mentioned in Section~\ref{subsec:inner_binary} the maximum orbital expansion of the inner binary ($a_{\rm in}$), which occurs after the SN explosion of the primary, is $a_{\rm in} = 766.5 \rsun$. With the latter,  $a_{\rm out} = 2300 \rsun$ and the evolution of the HTS can be modeled. The main stages are next listed:
\begin{enumerate}[i]
    \item At ZAMS, the tertiary star, of mass $M_{3} \simeq 35 \msun$ and metallicity $z = 0.0001$, is located in a circular orbit at $a_{\rm out} \sim 1817~\rsun$.
    \item Approximately 3.4 Myr after ZAMS, the primary star reaches RLOF and transfers $\Delta M \sim 67.1 \msun$, the secondary accretes $\sim 33.4\ \msun$ (the rest is lost). At this stage the tertiary could accrete a fraction of the mass which is lost (if its orbital velocity is comparable to the ejected mass velocity), but our model does not account for this. In response to the mass loss the binary orbit expands to $a_{\rm in} \simeq 737.1\ \rsun$, and the orbit of the tertiary star to $a_{\rm out} \simeq 2216\ \rsun$.  
    \item At 3.6 Myr, the primary star collapses onto a $45.8\ \msun$ BH (losing $6.1\ \msun$). The inner binary orbit expands to $a_{\rm in} = 766\ \rsun$, and the orbit of the tertiary star reaches the minimum stable orbit for the HTS with $a_{\rm out} = 2300\ \rsun$.
    \item At 4.6 Myr, the secondary star expands and produces a CE with the first-born BH. We assume that no accretion onto the BH occurs (although a few $\msun$ could be accreted); instead, $68\ \msun$ are ejected (not necessarily unbound) from the secondary star, which gets stripped down to $44.1\ \msun$. The ejected material, has low velocity and thus a large fraction of it may be captured by the tertiary star, i.e., it undergoes ceRLOF mass transfer. 
    \item During the CE phase of the binary system, the orbit of the inner binary decays to $a_{\rm in}=6.4\  \rsun$, while the tertiary star captures up to $68\ \msun$ of the ejected mass. Assuming conservative mass transfer during this ceRLOF phase, the mass of the tertiary star grows to $M_{3} \simeq 103 \msun$ which brings the mass ratio of the tertiary to the inner binary close to one ($q_3=M_{3}/(M_{1}+M_{2})=103/(45.8+44.1)=1.15$), thus reducing its orbit to $a_{\rm out} \simeq 818\ \rsun$. 
    \item At 4.9 Myr, the secondary star collapses to a $39.6\ \msun$ BH, losing $4.4\ \msun$ in the process. Both orbits expand a little after the mass loss to $a_{\rm in} \simeq 6.7\ \rsun$ and $a_{\rm out} \simeq 836\ \rsun$. The RL radius of the tertiary star reaches $R_{\rm RL,3} \simeq 330\ \rsun$.
    \item A few hundred kyrs later, the tertiary star ends its MS stage, expands, fills its $R_{\rm RL,3}$ and produces a CEE with the BBH. In this CEE, the BHs accrete several tens of solar masses, ending at $M_{\rm BH1} \simeq 86\ \msun$ and $M_{\rm BH2} \simeq 66\ \msun$. The CEE further decreases the inner orbital separation.  While, the loss of mass from the tertiary onto the binary decreases the mass ratio and expands the orbit of the tertiary to $a_{\rm out} > 2074\ \rsun$.
    \item Finally, the BBH will merge via GW emission in $\sim$ Myr. This timescale is similar to that when the tertiary star will become a BH, but a sequential BBH merger will not occur within the age of the Universe. Depending on what happens first, and the magnitude and direction of the recoil kick of the BBH product, the system might be disrupted or remain bound, now as a binary.
\end{enumerate}

\subsection{Star-forming region capture and CEE - BBH captures a massive star}
\label{subsec:ce-bbh-capture}

We consider the binary system described in Section~\ref{subsec:inner_binary}. If it captures a tertiary massive star companion ($M_3 > 35\ \msun$) before the CE phase, then we can obtain a HTS system which can also produce the desired BBH in the PISN. The tertiary star could capture $68~\msun$ of the binary system in a ceRLOF and evolve similarly to the HTS in Section~\ref{subsec:triple}. Given the CEE stage of the inner binary, the tertiary star could be captured into a much smaller orbital separation ($a_{\rm out} \gtrsim 30~\rsun$) without disturbing the evolution of the BBH. If this is the case, the CEE in which mass is transferred onto the BBH may occur in a shorter timescale than in our HTS scenario (as the tertiary star will RLOF at an earlier time).

Alternatively, the tertiary could be captured at a later stage (e.g. once the BBH has formed).  In this scenario the captured star must be more massive ($M_3 > 100 \msun$) given that it will not necessarily be able to capture mass during the ceRLOF of the binary. A caveat here would be that the ejected mass (the $68~\msun$) may not have escape velocity and fall back onto the binary (and be accreted by the BBH), and another part may form a circumbinary disk around the tertiary star. Eventually the captured tertiary would star evolve, expand, fills its RL and produce a CEE with the BBH where the BHs reach their final masses before merging ($M_1 > 86~\msun$ and $M_2 > 66~\msun$).

The capture of a wandering star, or of a star ejected from another binary/tertiary system may happen at different evolutionary phases of the binary. At early stages of BBH formation (e.g., stage {\it i} in Section~\ref{subsec:inner_binary}), the binary increases its mass (to $\approx 200~\msun$) and has a large cross section; however, the evolutionary lifetime as a stellar binary is relatively short ($\sim$ Myr). If the tertiary is captured during the next 200 kyr (stages {\it ii} and {\it iii}), given the decrease in total mass ($\sim 160~\msun$) and the increase in orbital separation ($\lesssim 766~\rsun$), the BBH will be more prone to disruption. Finally, after the CE phase (stage {\it iv}) the binary system will have a much smaller cross section ($a_{\rm in} \lesssim 7~\rsun$), less mass ($M_1 \gtrsim 46~\msun$ and $M_2 \gtrsim 40~\msun$), but will have a large amount of time available to capture a tertiary before merging ($\tau_{\rm GW} \sim 7$ Gyr). The expected capture rate is larger during the ZAMS phase, with twice the mass and 70 times larger orbital separation (so that the the cross section is $\sim 10^4$ larger) with respect to the BBH phase even though the timescale is 2000 times smaller. Also, the binary will be closer to the star formation region where it was formed, thus the stellar density will be larger.

\begin{table*}
\caption{PISN-BH candidates from \citet{abbott20c}. The columns correspond to the GW event, the mass of the BH remnant ($M_{\rm  rem}$), the masses of the merging BHs ($M_{\rm BH,1}$ and $M_{\rm BH,2}$), and the effective spin ($\chi_{\rm eff}$).}
\centering
\begin{tabular}{ |l|cccc| } 
\hline\hline
GW event & M$_{\rm rem}$ [$\msun$] & M$_{\rm BH,1}$ [$\msun$] & M$_{\rm BH,2}$ [$\msun$] & $\chi_{\rm eff}$ \\
\hline\hline
\rowcolor{lightgray}
GW190403\_051519 & $105.2^{+29.1}_{-24.1}$ & $88.0^{+28.2}_{-32.9}$ & $22.1^{+23.8}_{-9.0}$ & $0.70^{+0.15}_{-0.27}$\\ 
GW190426\_190642 & $175.0^{+39.4}_{-34.3}$ & $106.9^{+41.6}_{-25.2}$ & $76.6^{+26.2}_{-33.6}$ & $0.19^{+0.43}_{-0.40}$\\ 
\rowcolor{lightgray}
GW190519\_153544 & $101.0^{+12.4}_{-13.8}$ & $66.0^{+10.7}_{-12.0}$ & $40.5^{+11.0}_{-11.1}$ & $0.31^{+0.20}_{-0.22}$\\ 
GW190521         & $156.3^{+36.8}_{-22.4}$ & $95.3^{+28.7}_{-18.9}$ & $69.0^{+22.7}_{-23.1}$ & $0.03^{+0.32}_{-0.39}$\\ 
\rowcolor{lightgray}
GW190602\_175927 & $110.9^{+17.7}_{-14.9}$ & $69.1^{+15.7}_{-13.0}$ & $47.8^{+14.3}_{-17.4}$ & $0.07^{+0.25}_{-0.24}$\\ 
GW190620\_030421 & $87.2^{+16.8}_{-12.1}$ & $57.1^{+16.0}_{-12.7}$ & $35.5^{+12.2}_{-12.3}$ & $0.33^{+0.22}_{-0.25}$\\ 
\rowcolor{lightgray}
GW190701\_203306 & $90.2^{+11.3}_{-8.9}$ & $53.9^{+11.8}_{-8.0}$ & $40.8^{+8.7}_{-12.0}$ &$-0.07^{+0.23}_{-0.29}$\\ 
GW190706\_222641 & $99.0^{+18.3}_{-13.5}$ & $67.0^{+14.7}_{-16.2}$ & $38.2^{+14.6}_{-13.3}$ & $0.28^{+0.26}_{-0.29}$\\ 
\rowcolor{lightgray}
GW190929\_012149 & $101.5^{+33.6}_{-25.3}$ & $80.8^{+33.0}_{-33.2}$ & $24.1^{+19.3}_{-10.6}$ & $0.01^{+0.34}_{-0.33}$\\ 
GW191109 010717 & $107^{+18}_{-15}$ & $65^{+11}_{-11}$ & $47^{+15}_{-13}$ &  $-0.29^{+0.42}_{-0.31}$\\ 
\rowcolor{lightgray}
GW191127\_050227 & $76^{+39}_{-21}$ & $53^{+47}_{-20}$ & $24^{+17}_{-14}$ &  $0.18^{+0.34}_{-0.36}$\\ 
GW200220\_061928 & $141^{+51}_{-31}$ & $87^{+40}_{-23}$ & $61^{+26}_{-25}$ & $0.06^{+0.40}_{-0.38}$\\ 
\hline\hline
\end{tabular}
\label{tab2}
\end{table*}

\section{Discussion}
\label{sec:disc}

\subsection{Channels to produce BHs with masses in the PISN mass gap}

In this paper we propose two evolutionary channels in which, through HCA onto a BBH during the CE phase, BHs with masses falling in the theoretically predicted PISN mass gap may be produced. Thus, the BBH system progenitor of the GW source GW190521 may have been formed via such scenarios. 

Our results are based on two key processes: i) a ceRLOF, where a fraction of the material lost by the stellar progenitors of the BBH system is stored into a tertiary, massive star, and ii) HCA, which allows a large amount of mass to be accreted by the BBH, growing one of the BHs (or the two BHs) well above the PISN mass gap. We stress that, while the Eddington limit applies for spherical symmetry and assumes electron scattering for the momentum deposition by the photons in the accreting material, during HCA most of the energy is lost through neutrinos (whose mean free path is many orders of magnitude larger than that of photons). 

In the first channel, a binary system with $\sim 120~\msun$ and $\sim 80~\msun$ ZAMS stars evolves inside a HTS producing a BBH system of $\sim 45~\msun$ and $\sim 40~\msun$. In this scenario, most of the mass lost by the binary system is stored in a tertiary star which later returns it to the BBH enhancing both BH masses. In the second scenario, a tertiary star is captured by the binary system (which in turn evolves in a similar path as the first proposed evolutionary channel).

In both scenarios, the tertiary star evolves out of the MS stage after the BBH with BH masses of $M_1 \simeq{45}~\msun$ and $M_2 \simeq{40}~\msun$ has formed. The tertiary expands and engulfs the BBH in a CE, transferring through HCA in the CE phase $\sim 40 \msun$ to the $45-\msun$ BH, and $\sim 25 \msun$ to the $40-\msun$ BH, leaving behind two BHs with masses of $M_1 \simeq{85}~\msun$ and $M_2 \simeq{66}~\msun$. Then, a BBH system with an orbital separation $a_{\rm BS}\lesssim 10\; \rsun$ will merge in a timescale of $\tau_{\rm GW} \lesssim T_H/2$, where $T_H$ is the Hubble timescale\footnote{Using $z = 0.82\rm $ in \citet{2006FoPhL..19..277C} one obtains a light travel time $\lesssim 28/[1+\rm (1+z)^2]\; {\rm Gyr} = 6.5$ Gyr (with which the maximum time for the BBH to merge is $\approx 7.3$ Gyr).
From \citet[][]{misner17} the maximum orbital separation is:
  \begin{eqnarray}
    a &=& \left(\frac{256 G^3 \eta M^3 \tau_{GW}}{5 c^5}\right)^{1/4} \nonumber\\
      &\simeq& 55 \ \rsun  \left(\frac{\tau_{GW}}{7 {\rm Gyr}}\right)^{1/4} \left(\frac{M_{\rm BBH}}{150\msun}\right)^{3/4}   \left(\frac{\eta}{0.25}\right)^{1/4} 
  \end{eqnarray}
  with $M_{\rm BBH} = M_{\rm BH,1} + M_{\rm BH,2} = 85\ \msun + 65\ \msun = 150\ \msun$, $\eta = (M_{\rm BH,1} M_{\rm BH,2})/M_{\rm BBH}^2 \simeq 0.25$. This corresponds to an orbital period of 3.88 d.} in a galaxy which is $5.3$ Gpc ($z = 0.82$) from us.

\subsection{Possible progenitors}
\label{sec:obs}

MelnicK 34 (Mk34) is a spectroscopic binary, X-ray colliding-wind system, with two hydrogen rich WR stars \citep{crowther2011, hainich2014}. Mk34 has a $155.1 \pm 1$ d orbital period, and a projected distance of 2 pc from R136 (the stellar cluster with the most massive stars in 30 Doradus,  \citealt{pollock2018}). Its current age is $\sim0.6\pm0.3$ Myr, and the masses of the WRs are $139^{+21}_{-18}~\msun$ for the primary  and $127^{+17}_{-17}~\msun$ for the secondary, corresponding to ZAMS masses of $144^{+22}_{-17}~\msun$ and  $131^{+18}_{-16}~\msun$ respectively \citep{2019Tehrani}. Thus, this system represents a possible progenitor of the BBH studied in this paper.

Modeling the future evolution of Mk34, the primary star will have a maximum radius of $R_{\rm max} \sim 78-126~\rsun$ during MS, with a periastron orbital separation of $254~\rsun$ \citep{2019Tehrani}. Also, the primary star in Mk34 will fill its Roche lobe (with $400 - 500~\rsun$) and begin a mass transfer phase before it forms a BBH system (with masses $<45~\msun$) in 2 to 2.5 Myr \citep{yusof13}. A more recent study using COMPAS finds that the future evolution of Mk34 is uncertain \citet{2022Belczynski}, allowing for a system like the one predicted in Section~\ref{subsec:inner_binary}. Thus, it is possible that even at metallicities as large as those of LMC these kind of BBH mergers may occur. If Mk34 was accompanied by a $35~\msun$ tertiary or if the binary system captures a tertiary star (probably another runaway star from R136) within the mass range $50~\msun \lesssim M_3 \lesssim 100~\msun$ and orbital separation as large as $\sim$ 10~AU, then the system would follow a path similar to that described in Section~\ref{sec:scenarios}. HCA in the CE phase could lead to BBH system with masses very close to those from GW190521.

GW190521 is also the first BBH merger which may be associated to an electromagnetic (EM) counterpart. The optical flare (with $E\sim10^{51}$ erg) could be interpreted as evidence of interaction between the BBH and the accretion disk of an active galactic nucleus \citep{graham20}. However, GW190521 was at a redshift $\simeq 0.82^{+0.28}_{-0.34}$ \citep{abbott20b}, and the EM event was detected $\sim 34$ d after the GW signal. Thus, the association between GW190521 and the EM radiation is weak and remains to be confirmed. Nonetheless, we must note that, in our scenarios, if the BBH were to merge during the the CE of the tertiary star, it is likely that an energetic EM counterpart (GRB-like, in terms of energy) would be observed at the time of the merger \citep[similar to the scenarios discussed by][]{2021MNRAS.506.2445S}. This would be particularly the case if the BBH merger occurs in the core of the star. The BBH merger, unlike a binary neutron star merger, produces its jets from accretion disks formed from external material, whereas binary neutron star may produce their own disk from neutron star material \citep{Lattimer1976, Rosswog1999, Metzger2011}. The amount of angular momentum available for the infalling material is high, so, most likely a pair of extremely energetic jets would form and produce an energetic electromagnetic transient counterpart to the GW signal. Fallback from the exploded star could also help to produce a late EM signal (a hypernova-like transient).

CEE of a BBH in a tertiary star could result in mergers within the CE if the drag is strong. This can be prevented if the BHs accrete at BHL rates. 
However, a merger of the BBH inside the core of the triple star can also account for a GW signal accompanied by a GRB.

\subsection{BH spins}

Table~\ref{tab2} shows a list of GW events which have at least one of the merging BHs in the PISN mass gap \citep{abbott20c}. These BHs could be produced by the HCA onto the BBH system during the CE of the tertiary star (see Section~\ref{sec:scenarios} for further details). Apart from the formation channels which may produce massive BHs, we analyse the spin that such BHs would have. 

Assuming that during the formation of the BH (or BHs) in the PISN mass gap there are weak SN kicks, mass transfer stages, and tidal synchronization episodes, the expectation is that massive BHs will be formed with low spins which would tend to be aligned with the orbital angular momentum (see, e.g., discussion in \citealt{2008MMBLP,2011MM}; unless something such as SN kicks may alter them \citealt{2022Tauris}). We highlight the fact that the effective spins $\chi_{\rm eff}$ are distributed over a large range of values (from $-0.29$ to $0.70$) and that there is no strong trend. 
Values of $\chi_{\rm eff}\approx 0.70$ are expected in BBHs which experience multiple sequential mergers. Focusing in our triple system evolutionary channels, we enumerate a few mechanisms that will likely increase and/or misalign the spins of the BHs (and which may be taken into account when predicting the characteristics of the resulting BHs since massive stars are preferentially in multiple stellar systems, see e.g. \citealt{2017ApJS..230...15M}):
\begin{enumerate}
    \item Strong SN kicks can misalign as well as spin up BHs.
    \item HCA during CE phases, especially when the BH mass doubles, can turn a spinless BH to a Kerr BH (see Figure~6 in \citealt{2000Brown}) and align the BH spin to the orbital angular momentum of the binary system.
    \item HCA in a CE of the tertiary star where the orbits are misaligned could produce almost anything, from aligned small spins to completely misaligned large spins.
    \item Tertiary-star orbital plane misaligned with binary orbital plane (particularly likely in captures) can misalign the BH spins (Kozai-Lidov).
\end{enumerate}

\subsection{Other considerations}

The amount of mass accreted during the CE stage, estimated in Section~\ref{sec:scenarios} should be taken as an upper limit. The error bars of the BH masses in GW190521 are $\sim 15-20$\% at $1\sigma$ \citep{abbott20a}. So, it is possible that the BHs may have lower masses, thus reducing the amount of mass accreted. The PISN mass gap for BHs ($M_{\rm GAP}$) is limited between $55 \lesssim M_{\rm GAP}/\msun \lesssim 135$ \citep{2001FWH} but the lower limit may be higher. The latter may increase to $65\ \msun$ considering low mass-loss during the evolution of massive ZAMS stars \citep{woosley2017, wh2021}. Other studies propose that it could be of order 80 or 90 $\msun$ \citep{2020Belczynski, Belczynski2010}. Thus, by taking an upper BH mass limit $\sim 65\ \msun$, our models can form a BBH with M$_{\rm BBH,1} \simeq 65\ \msun$ and M$_{\rm BBH,2} \simeq 60\ \msun$, hence reducing the mass accreted by $\sim$ a half (to $\sim 25$ M$_\odot$) in order to form the observed BBH in GW190521.

Stellar models which attempt to form BHs with masses within the PISN mass gap, assume that the hydrogen envelope collapses onto the BHs \citep[e.g.,][]{renzo20}, and drives them directly into the PISN; after that, they still need to find another BH which goes through a similar process to merge with. In our model we get BHs with masses above the PISN mass gap via the ceRLOF mechanism by transferring the hydrogen envelope onto the tertiary star and returning it to the BBH at a later stage.

Another mechanism which could lead a stellar binary to evolve into a GW190521-like system is when a star with $M_\star \gtrsim 100~\msun$ passes close to the BBH and is tidally disrupted (or a Mk34-like system with $\sim 50-\msun$ BHs is produced). In this case the BHs accrete several tens of solar masses of disrupted-star material, bringing them into the mass range of GW190521 as well as closer together. Mass loss in pulsational PISNe may also lead to mass transfer onto the tertiary star, especially for material which does not acquire escape velocity or with a velocity $v \sim v_{\rm orb}$. This material is usually lost during single stellar evolution, but a fraction of it can be recovered in binary- or triple-system evolution.

The presence of strong jets and/or large $\alpha\lambda$ values could prevent efficient mass transfer during ceRLOF. The formation of jets/winds/outflows from accretion disks around the accreting BHs can remove the envelope of a massive star, thus they can limit the mass growth of the BHs. A small orbital separation between the BBH components can limit the formation of large accretion disks. Thus the orbital separation of the BBH could be an important factor for the amount of mass accreted by the BHs. Another relevant factor is that the ore massive the BHs the larger the BHL radius, thus it is likely that substantial mass growth by HCA during a CE will also depend on the initial mass of the BHs as they undergo CEE.

\citet{2021Klencki} argue that unless the CEE occurs when the envelopes are mostly convective, these cannot be ejected. Then, assuming that accretion is negligible, the CE mass which is not ejected produces an orbital drag, and the binary merges. In our case, the mass which is not ejected ends up being rapidly accreted by the BHs, thus, preventing a merger.
Also, in a triple system, mass can be removed from the inner binary by reaching the Roche lobe, without having to reach the escape velocity. It is then transferred onto the (tertiary) companion. Hence, the closer the tertiary star is, the more efficient the ceRLOF phase.

\section{Conclusion}
\label{sec:concl}

In this paper we propose an evolutionary channel in which a BBH system, via HCA during the CE phase, produces BHs with masses within the PISN mass gap (e.g., the BBH system which produced the GW190521 event). We have considered two main scenarios. In the first scenario, the evolution of a HTS with an inner binary system with two very-massive stars. During the evolution of these two stars, part of the material is shed away and it is stored in a tertiary companion  (our {\it ceRLOF} mass transfer mode as described in Section \ref{sec:MTnMS}). The massive binary produces a BBH system with masses below the PISN gap limit (e.g., $\sim$ 45 M$_{\sun}$ and $\sim$ 40 M$_{\sun}$). Later, the tertiary star leaves the MS, expands and engulfs the BBH in its envelope, the BBH accretes back the previously lost mass and results in a BBH with masses in the PISN mass gap.
In the second scenario, a massive binary captures a tertiary star. If the star is captured during the MS of the binary, it may capture by ceRLOF mass from the inner binary and proceed similarly to the first model. Alternatively, the captured star may be a very-massive star, captured after the BBH has formed.  As the tertiary evolves out of its MS stage it produces a CE with the BBH and transfers mass onto it, leaving, again a BBH with masses in the PISN mass gap.

The main bottleneck for these scenarios may come from whether the BBHs are faster at accreting the CE material than the CE at absorbing energy from the system and unbounding it. If these processes are nearly equally efficient, the accretion during a CE stage of a star with $M_\star \gtrsim 100 \msun$ by the BBH could account for the masses of the GW190521 event.

We also discuss possible progenitors of BBH systems with massive components. Mk34 represents a possible progenitor of the BBH studied in this paper. BBHs which are born in multiple systems will produces massive BHs with a disperse distribution of $\chi_{\rm eff}$ values.

Our paper takes simulations of binaries from a population-synthesis code (COMPAS) and includes them in the simplified analytical evolution of a HTS. Future work should study the hydrodynamic evolution of the ceRLOF, the role of HCA in BBHs, and incorporate them into a triple-star population-synthesis software. Such population study would determine rates and configurations of BBHs in the PISN mass gap via the ceRLOF+HCA formation mechanism described in this manuscript. The rates and configurations could be then compared with the current GW catalog and the forthcoming detections of current and future GW observatories.

\section*{Acknowledgements}

F.D.C and D.L.C. acknowledge support from the UNAM-PAPIIT grant AG100820.
D.L.C. is supported by C\'atedras CONACyT at the Instituto de Astronom\'ia (UNAM).
A.V.G. acknowledges support by the Danish National Research Foundation (DNRF132).
This research has made use of NASA’s Astrophysics Data System as well as arXiv.



\end{document}